\documentclass[sigconf]{acmart}
\usepackage{multirow}
\usepackage{graphicx}
\usepackage{subcaption}
\usepackage{enumitem}
\usepackage{adjustbox}
\usepackage{tabularx}
\usepackage{balance}
\AtBeginDocument{%
  }

\copyrightyear{2025}
\acmYear{2025}
\setcopyright{acmlicensed}
\acmConference[CIKM '25] {Proceedings of the 34th ACM International Conference on Information and Knowledge Management}{November 10--14, 2025}{Seoul, Republic of Korea}
\acmBooktitle{Proceedings of the 34th ACM International Conference on Information and Knowledge Management (CIKM '25), November 10--14, 2025, Seoul, Republic of Korea}
\acmISBN{979-8-4007-2040-6/2025/11}
\acmDOI{10.1145/3746252.3761248}

\begin{document}
\setlist[itemize]{left=0pt}
\title{Generative Data Augmentation in Graph Contrastive Learning for Recommendation}

\author{Yansong Wang}
\affiliation{%
  \institution{Southwest University}
  \department{College of Computer and Information Science}
  \city{Chongqing}
  \country{China}}
\email{yansong0682@email.swu.edu.cn}

\author{Qihui Lin}
\affiliation{%
  \institution{Southwest University}
  \department{College of Computer and Information Science}
  \city{Chongqing}
  \country{China}}
\email{linqihui@email.swu.edu.cn}

\author{Junjie Huang}
\affiliation{%
  \institution{Southwest University}
  \department{College of Computer and Information Science}
  \city{Chongqing}
  \country{China}}
\email{junjiehuang.cs@outlook.com}

\author{Tao Jia}
\authornote{Corresponding author.}
\affiliation{%
  \institution{Southwest University}
  \department{College of Computer and Information Science}
  \city{Chongqing}
  \country{China}
}
\affiliation{
  \institution{Chongqing Normal University}
  \department{College of Computer and Information Science}
  \city{Chongqing}
  \country{China}
}
\email{tjia@swu.edu.cn}

\renewcommand{\shortauthors}{Yansong Wang, Qihui Lin, Junjie Huang, and Tao Jia}

\begin{abstract}
Recommendation systems have become indispensable in various online platforms, from e-commerce to streaming services. A fundamental challenge in this domain is learning effective embeddings from sparse user-item interactions. While contrastive learning has recently emerged as a promising solution to this issue, generating augmented views for contrastive learning through most existing random data augmentation methods often leads to the alteration of original semantic information. In this paper, we propose a novel framework, GDA4Rec (\textbf{G}enerative \textbf{D}ata \textbf{A}ugmentation in graph contrastive learning for \textbf{Rec}ommendation) to generate high-quality augmented views and provide robust self-supervised signals. Specifically, we employ a noise generation module that leverages deep generative models to approximate the distribution of original data for data augmentation. Additionally, GDA4Rec further extracts an item complement matrix to characterize the latent correlations between items and provide additional self-supervised signals. Lastly, a joint objective that integrates recommendation, data augmentation and contrastive learning is used to enforce the model to learn more effective and informative embeddings. Extensive experiments are conducted on three public datasets to demonstrate the superiority of the model. The code is available at: \url{https://github.com/MrYansong/GDA4Rec}.
\end{abstract}



\maketitle

\section{Introduction}
\label{sec:introduction}
The recommendation system plays an increasingly important role in our digital society, providing items that a user is truly interested in from a vast of potential candidates \cite{cheng2016wide, he2017neural, zhang2019deep}. A recommendation system usually makes suggestions based on the embedding of the user and item. Such information can be obtained through the temporal pattern in a user's item click sequence \cite{fan2021lighter}, the association among a user's online actions (such as viewing the item, adding the item to the shopping cart, and so on) \cite{wu2022multi}, the relationship among different items \cite{chen2023heterogeneous}, and more. 

Inspired by the capability of Graph Neural Networks (GNNs) in encoding relationships among nodes in a network, GNN-based recommendation models are proposed that demonstrate great potential \cite{he2020lightgcn}. However, the record of user-item interaction is usually sparse, leading to less effective performance or the risk of overfitting under the framework of GNN. To cope with the few-shot learning problem in recommendation, contrastive learning is adopted and becomes a very effective approach. Contrastive learning incorporates self-supervision signals from data augmentation without requiring additional manually annotated labels. The data augmentation can be done by perturbing the original graph with uniform node/edge dropout or random feature masking to synthesize supplementary views \cite{velivckovic2018deep, peng2020graph, bachman2019learning}. Unfortunately, altering the topological structure, especially in sparse graphs, may directly change the semantics of nodes, giving rise to the bias in the learned embedding. An alternative approach is to sample a subset of nodes and corresponding edges to generate a subgraph as the augmented view, using techniques such as uniform sampling \cite{shuai2022review}, ego-net sampling \cite{cao2021bipartite}, knowledge-based sampling \cite{yu2021self} and so on. The effectiveness of this approach relies on the choice of the subgraph samples. Given the diverse network topology, a sampling method effective in one network may not work in another network. The method is hard to be generalized.

\begin{figure}[htbp]
\centering
\includegraphics[width=0.45\textwidth]{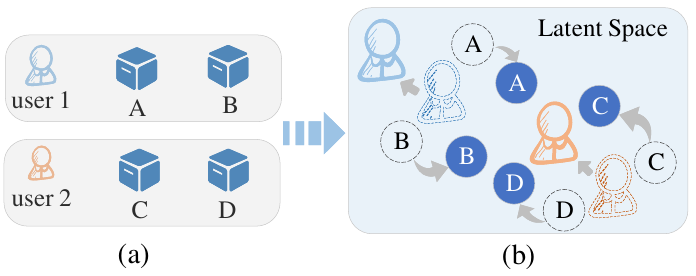}
\hfill
\caption{(a) Interaction records between user1 and user2. (b) Embeddings of users and items in the latent space, where dashed and solid lines denote their positions in the original and augmented views, respectively.}
\label{fig:random_perturbation}
\end{figure}

Recent advancements \cite{yu2022graph, yu2023xsimgcl, zhu2024distribution, ouyang2024improve} have introduced embedding-level data augmentation strategies that enhance representational diversity by incorporating controlled perturbations within the embedding space. These methods carefully inject constrained noise into node embeddings, ensuring that the underlying semantics and structural properties of the original graph are retained within the augmented data. When integrated with contrastive learning frameworks, the synergy between data augmentation and contrastive objectives facilitates more robust feature embeddings, thereby enhancing the model's capability to capture complex user-item interactions in recommendation tasks \cite{wu2021self, yu2022graph}. These strategies seek to maintain invariance between the original and augmented perspectives by applying controlled perturbations, but the random noise drawn from some simple probability distribution may inadvertently introduce extraneous features that are not inherently aligned with the original data \cite{zhu2024distribution, ouyang2024improve}. This can result in a divergence in the augmented view from the original feature distribution. As shown in Figure \ref{fig:random_perturbation}, the randomly perturbed embeddings of users and items corrupt the semantic information inherent in the original interaction records, causing embeddings of items $A$ and $B$ being greater similarity to that of user2.
Concurrently, during the training process, the embedding of users and items should progressively align with their true underlying distribution. The fixed-scale application of random noise poses challenges in accurately reflecting the intrinsic variability of the data.
Consequently, developing a contrastive recommendation framework that incorporates adaptive noise scales based on the characteristics of users and items remains imperative.

In this work, to solve the challenges mentioned above, we propose a novel and generic recommendation framework, namely GDA4Rec (\textbf{G}enerative \textbf{D}ata \textbf{A}ugmentation in graph contrastive learning for \textbf{Rec}ommendation). Particularly, we employ a generative model to extract augmented views from the latent space of the data with two loss functions. The first one is reconstruction loss, which ensures the generative noise closely resembles the original data, while the other is distribution discrepancy loss, which promotes distribution coherence within the latent space. This dual-loss effectively facilitates the generation of realistic and diverse augmented views. Besides, we consider that items with strong complementarity are easily interacted with by the same user, and then extract the complement matrix from the user-item interaction to generate more self-supervision signals. 
The major contributions are summarized as follows:
\begin{itemize}
\item  We introduce a new generic framework for contrastive recommendation tasks, designed to construct more reliable augmented views through the deep generative model to strengthen the self-supervision signal of the model.

\item  In data augmentation, we generate noise from the user and item embeddings using a generative model rather than perturbing the original embeddings with fixed-scale random noise. This strategy enables the creation of augmented views adaptively while preserving similarity with the original feature distribution.

\item To strengthen the supervised signal, we derive an item complement matrix from user-item interactions and introduce a filtering mechanism to eliminate redundant information. This strategy incorporates item complementarity into recommendations.

\item We conduct experiments on three real-world datasets to validate the effectiveness of our model, complemented by ablation studies that demonstrate the contribution of each module.

\end{itemize}

\section{Related Work}
\subsection{GNN Based Recommendation}
The core mechanism of recommendation systems lies in leveraging the similarities between users and items to predict potential user preferences. Early recommendation models based on representation learning primarily involved techniques such as matrix factorization \cite{koren2009matrix} and factorization machines \cite{rendle2011fast}. 
Their performance is often constrained when dealing with high-dimensional and complex datasets. With advances in computational power, neural network based models \cite{li2015deep, he2017neural} have significantly improved predictive performance. Recently, graph neural networks have emerged as powerful approaches in recommendation systems \cite{ying2018graph, wang2019neural, he2020lightgcn, gao2023survey}, which leverages graph structures to model high-order connectivity among users, items, and their interactions. NGCF \cite{wang2019neural} proposes a spatial GNN that integrates the high-order connectivity into the embedding process by propagating embeddings on the graph. LightGCN \cite{he2020lightgcn} simplifies traditional GCNs by eliminating feature transformation and nonlinear activation functions. DCAN \cite{wang2022self} proposes a new two-channel attention network to leverage sequential information and complex project transformations. Additionally, there are other models, such as HGCL \cite{chen2023heterogeneous}, that focus on modeling user-user and item-item relationships for graph-based collaborative filtering.

\subsection{Contrastive Learning for Recommendation}
Contrastive learning is a type of self-supervised learning that enables models to learn evenly distribution of users and items \cite{wu2021self}. 
In the context of collaborative filtering recommendation, SGL \cite{wu2021self} introduces a contrastive learning based model that utilizes traditional graph data augmentation methods to perturb the original graph structure. BUIR \cite{lee2021bootstrapping} introduces augmented views based on the neighborhood information of each user and item. 
SimGCL \cite{yu2022graph} challenges traditional graph augmentation methods by adding uniform noise to the embeddings of GNN layers. 
NCL \cite{lin2022improving}, inspired by \cite{lin2022prototypical}, identifies the latent prototype of a user through clustering to explicitly capture the potential node correlation in contrastive learning.
Besides these model architecture studies, some works focus on studying the influence of important modules in contrastive learning. For example, \cite{wu2021rethinking} explores the impact of the number of negative samples on InfoNCE \cite{oord2018representation} loss and proposes an adaptive negative sampling strategy.

\subsection{Data Augmentation in Contrastive Learning}
Recent research has shown that contrastive learning critically relies on contrastive views \cite{jing2023contrastive}. However, most models do not naturally have multiple views as input. Thus data augmentation is needed to generate contrastive views. Graph-based augmentations are the most widely applied, including techniques such as node dropout \cite{wu2021self}, edge dropout \cite{yang2022knowledge}, graph diffusion \cite{long2021social}, and subgraph sampling \cite{you2020graph}. Such augmentation typically exhibits limited generalization in recommendation models, since different topologies require distinct augmentation strategies.
Model-based augmentations such as message dropout \cite{wang2023cl4ctr}, embedding noise \cite{yu2022graph}, and parameter noise \cite{xia2022simgrace}, demonstrate strong performance in addressing these challenges. SSL4Rec \cite{yao2021self} adopts the classical two-tower structure for contrastive learning and utilizes feature correlations for data augmentation. 
DAHNRec \cite{zhu2024distribution} proposes a distribution-aware method that generates suitable user and item representation to enhance the uniformity of the embedding space. Similarly, VGCL \cite{yang2023generative} employs a generative model to directly learn node embeddings from the interaction graph, while neglecting the primary recommendation task.
AUPlus \cite{ouyang2024improve} devises the 0-layer perturbation mechanism that augments the data for self-supervised contrastive learning to promote label-irrelevant alignment and uniformity.
Moreover, there are some works that explore entirely new augmentation techniques, such as singular value decomposition \cite{zhang2023spectral, cai2023lightgcl}. These methods are more general than graph-based augmentations and applicable to a variety of datasets.

\begin{figure*}[htp]
\centering
\includegraphics[width=\textwidth]{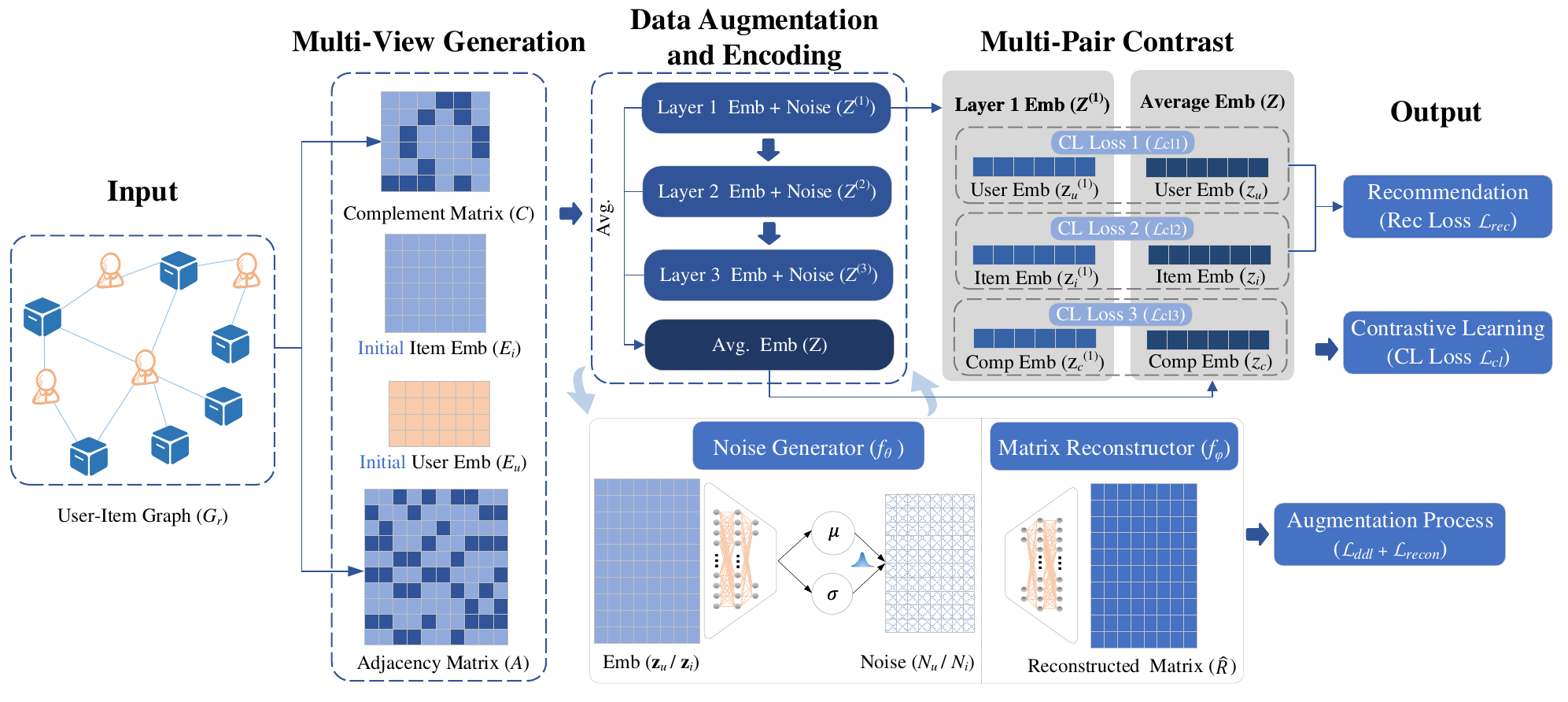}
\caption{The framework of the proposed model is shown above, which includes five parts: input, multi-view generation, data augmentation and encoding, multi-pair contrast and output.}
\label{framework}
\end{figure*}

\section{Proposed Framework}

\label{section: framework}
\subsection{Preliminaries}
For the recommendation, $G_{r}$ represents the user-item graph. Let $\mathcal{V}_u(|\mathcal{V}_u|=m)$ denote the user set and $\mathcal{V}_i(|\mathcal{V}_i|=n)$ denote the item set, which both come from the $G_{r}$. Considering user feedback, we use $R \in \mathbb{R} ^ {m \times n}$ obtained from the user-item graph $G_r$ to denote the user-item interaction matrix, where $R(u,i)=1$ if user $u$ has interacted with item $i$, otherwise $R(u,i)=0$.
Specifically, there are two kinds of matrices generated by $R$ used as inputs of the model encoder, including user-item adjacency matrix $A \in \mathbb{R} ^ {(m+n) \times (m+n)}$ and item complement matrix $C \in \mathbb{R} ^ {n \times n}$, where
\begin{equation}
    A =\bigg[\begin{matrix}
        \mathbf{0}^{m \times m} & R \\ R^T & \mathbf{0}^{n \times n} \\
    \end{matrix} \bigg].
\end{equation}
As for item complement matrix $C$, it comes from the interaction matrix $R$ that has undergone matrix transformation and been filtered, where each element in $C$ indicates how frequently two items are interacted by the same user. 

Furthermore, we denote randomly initialized user embedding and item embedding by $E_{u} \in \mathbb{R} ^ {m \times d}$ and $E_{i} \in \mathbb{R} ^ {n \times d}$ respectively, 
where $d$ is the embedding dimension we set.
Besides, ego embedding is also required, which is obtained by directly concatenating the user embedding and the item embedding, denoted by $E_{ui} \in \mathbb{R} ^ {(m+n) \times d}$.
Throughout the model, we use $\mathbf{z}$ to represent the final learned representations.

The objective of conventional embedding-based recommendation systems is to find the optimal $\mathbf{z}$ that maximizes its conditional probability,
\begin{equation} 
    \mathbf{z} = \mathop{\arg\max}\limits_{\mathbf{z}} P(Y| \mathbf{z}_u, \mathbf{z}_i),
\end{equation}
where $Y$ represents the interaction scores between users and items. In contrastive learning based methods, new augmented views $g$ are introduced via data augmentation to mitigate data sparsity,
\begin{subequations}
\begin{align}
    P(Y| \mathbf{z}_u, \mathbf{z}_i) &= \sum_{\mathcal{G}}P(Y| \mathbf{z}_u, \mathbf{z}_i, g)  P(g |\mathbf{z}_u, \mathbf{z}_i) \label{equ: original_1} \\
                   &= P(Y| \mathbf{z}_u, \mathbf{z}_i, f_g(\mathbf{z}_u, \mathbf{z}_i)),  \label{equ: original_2}
\end{align}
\end{subequations}
where $\mathcal{G}$ is the sample spaces of $g$. In particular, Eq. (\ref{equ: original_1}) follows the law of total probability. Since $g$ can only take a value $f_g(\mathbf{z}_u, \mathbf{z}_i)$ if $\mathbf{z}_u$ and $\mathbf{z}_i$ are given, the sum over $\mathcal{G}$ is removed, \textit{i.e.} $P(g |\mathbf{z}_u, \mathbf{z}_i) = 1$. Now, the problem is transformed into constructing $f_g(\mathbf{z}_u, \mathbf{z}_i)$. Conventional methods include perturbing the original graph, subgraph sampling, and so on. In this paper, to bridge the gap between original semantics and data augmentation, we generate augmented views through a deep generative model based on the $\mathbf{z}_u$ and $\mathbf{z}_i$, formulated as $f_g(\mathbf{z}_u, \mathbf{z}_i)$. And then we use a model $f(\cdot)$ to calculate the conditional probability $P(Y| \mathbf{z}_u, \mathbf{z}_i, f_g(\mathbf{z}_u, \mathbf{z}_i))$, similar to general recommendation systems.

\subsection{Methodology}
\subsubsection{Model Architecture}

The model is divided into five parts: input, multi-view generation, data augmentation and encoding, multi-pair contrast and output, as shown in Fig. \ref{framework}. 

The user-item graph $G_{r}$ containing interaction relationships between users and items is input as data sources.
Then in the multi-view generation phase, the user and item embeddings are generated by random initialization. Two matrices are generated by $G_{r}$ including the adjacency matrix $A$ and the complement matrix $C$.
In the data augmentation process, we use a single generator and reconstructor to generate the corresponding noise based on the embeddings at each layer.
For data encoding, We employ a multi-layer GNN to produce the final three representations, namely user embedding $\mathbf{z}_{u}$, item embedding $\mathbf{z}_{i}$, and complement embedding $\mathbf{z}_{c}$.
In the multi-pair contrast, three representations perform contrastive learning with corresponding contrastive views. A multi-task learning strategy \cite{jing2023contrastive} is then implemented to optimize the model parameters, and the final representations are used for ranking the candidate list.

\begin{figure}[tp]
\centering
\includegraphics[width=0.9\linewidth]{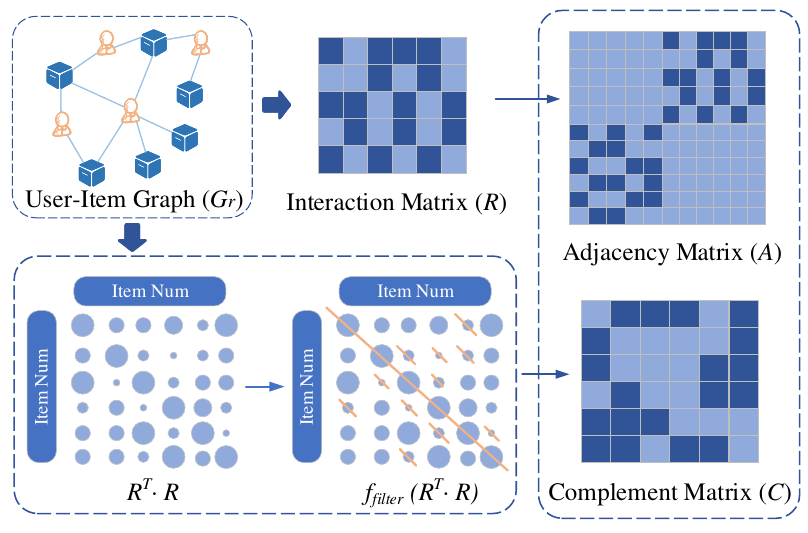}
\caption{The process of generating the adjacency matrix $A \in \mathbb{R} ^ {(m+n) \times (m+n)}$ and complement matrix $C \in \mathbb{R} ^ {n \times n}$ by the user-item graph $G_r$. $R \in \mathbb{R} ^ {m \times n}$ stands for interaction matrix.}
\label{view generation}
\end{figure}

\subsubsection{Multi-View Generation}
Generally, recommendation systems concentrate on user-item interactions while overlooking the underlying supplementary information behind these interactions. In this section, we omit the conventional step of constructing the adjacency matrix $A$ from the interaction graph $G_r$ and instead focus on building the item complement matrix $C$, the process is shown in Fig. \ref{view generation}.

Item complementarity refers to a relationship where two different items are tied and complement each other to fulfill a desire or need. To obtain the complement matrix of items, the common practice is as follows, 
\begin{equation}
   \hat{C}= R^{T} \cdot R,
\end{equation}
where $R \in \mathbb{R} ^ {m \times n}$ is the user-item interaction matrix, $R^{T}$ is its transpose, and $\hat{C} \in \mathbb{R} ^ {n \times n}$ is a original complement matrix. 
Each entry in the matrix $\hat{C}$ reflects how often a user has interacted with the corresponding pair of items. Larger values denote a higher degree of complementarity between the two items, making it more probable for a user to interact with them concurrently. 

The matrix $\hat{C}$ presents several issues that need to be addressed. First, the values in $\hat{C}$ vary significantly, with small values dominating the matrix. These low values contribute minimally and can introduce noise when calculating complementarity, so they should be filtered out.
Furthermore, we remove the self-loops inspired by LightGCN \cite{he2020lightgcn} and demonstrate its effectiveness through experiments. This operation is illustrated by the function $f_{\mathrm{filter}}(\cdot)$ in Fig. \ref{view generation}. Finally, we normalize $\hat{C}$ to obtain the final item complement matrix $C$, defined by the following formula,
\begin{equation}
   C= f_{\mathrm{norm}}(f_{\mathrm{filter}}(R^{T} \cdot R-diag(R^{T} \cdot R))),
\end{equation}
where $f_{\mathrm{norm}}(\cdot)$ is the normalization function, $f_{\mathrm{filter}}(\cdot)$ denotes the filtering function that assigns values below $\gamma$ to 0, $\gamma$ is a hyperparameter controlling the filtering threshold, $R^{T}$ is the transpose matrix of $R$, and $diag(\cdot)$ denotes the extracted diagonal matrix.

\subsubsection{Data Augmentation and Encoding}
Previous work has demonstrated the significant role of data augmentation ($f_g(\mathbf{z}_u, \mathbf{z}_i)$ in Eq. (\ref{equ: original_2})) in recommendation systems. To establish a connection between prediction and augmentation while ensuring that the augmented data closely resembles the original data, We introduce noise $N$ for data augmentation,
\begin{subequations} \small
\begin{align}
    &\log P(R| \mathbf{z}_u, \mathbf{z}_i) = \log \sum_{\mathcal{N}}P(R| N)  P(N |\mathbf{z}_u, \mathbf{z}_i) \label{equ:Z_1} \\
    &\geq \sum_{\mathcal{N}}Q(N)[\log P(R| N) + \log P(N | \mathbf{z}_u, \mathbf{z}_i) - \log Q(N)] \label{equ:Z_2} \\
    & = \mathbb{E}_{\mathcal{N} \sim Q(N)}[\log P(R| N) + \log \frac{P(N | \mathbf{z}_u, \mathbf{z}_i)}{Q(N)}], \label{equ:Z_3}  
\end{align}
\end{subequations}
where $\mathcal{N}$ is the sample spaces of $N$. In particular, Eq. (\ref{equ:Z_1}) follows the law of total probability. We introduce an approximate prior distribution $Q(N)$ and use Jensen inequality to get Eq. (\ref{equ:Z_2}) and Eq. (\ref{equ:Z_3}). To maximize $P(R| \mathbf{z}_u, \mathbf{z}_i)$, we design a generative noise module. This module takes $\mathbf{z}$ as input, uses a generator $f_\theta$ to produce noise $N$ that follows a specific distribution $Q(N)$, and uses a reconstructor $f_\phi$ to reconstruct the interaction matrix $R$. In this study, we adopt a Gaussian distribution as prior. 
Leveraging the idea from Variational AutoEncoders (VAE), we employ neural networks to generate the mean $\mu_\theta(\mathbf{z}_u, \mathbf{z}_i)$ and variance $\sigma_\theta^2(\mathbf{z}_u, \mathbf{z}_i)$, corresponding to $f_\theta(N | \mathbf{z}_u, \mathbf{z}_i) = \mathcal{N}(\mu_\theta(\mathbf{z}_u, \mathbf{z}_i), \sigma_\theta^2(\mathbf{z}_u, \mathbf{z}_i))$. While the random sampling approach preserves the randomness of the noise $N$, this process is non-differentiable. To address this, we apply the reparameterization trick \cite{kingma2013auto} to make the process differentiable,
\begin{equation}
    f_\theta (N | \mathbf{z}_u, \mathbf{z}_i) = \mu_\theta(\mathbf{z}_u, \mathbf{z}_i) + \sigma_\theta^2(\mathbf{z}_u, \mathbf{z}_i) \odot \epsilon,
\end{equation}
where $\epsilon \sim \mathcal{N}(0,I)$ is an auxiliary variable that follows a normal distribution, $\odot$ is element-wise multiplication. In this way, the sampling process only involves linear operations, which are differentiable.
The objective of maximizing $P(R| \mathbf{z}_u, \mathbf{z}_i)$ is transformed into,   
\begin{equation} \small
\begin{split}
    \min -\mathbb{E}_{\mathcal{N} \sim Q(N)} & \big( \log f_\phi(
    R| N) + \log \frac{Q(N)}{f_\theta (N | \mathbf{z}_u, \mathbf{z}_i)} \big) \\
    = \min -\mathbb{E}_{\mathcal{N} \sim Q(N)} & \log f_\phi (R| N) \\
    &+ KL[Q(N) || f_\theta (N | \mathbf{z}_u, \mathbf{z}_i)].
\end{split}
\label{equ: loss noise gene}
\end{equation}
In Equation (\ref{equ: loss noise gene}), the first term is the reconstruction error between the original interaction matrix and the reconstructed matrix. We use an inner product to compute the propensity score that node $i$ connected with node $j$,
\begin{equation}
    P(\hat{R}_{ij}|N_i, N_j) = N_i^T \cdot N_j.
\end{equation}
The strategy to minimize the reconstruction error is,
\begin{equation}
    \mathcal{L}_{recon} = \mathbb{E}[({\hat{R}_{ij} - R_{ij}})^2 | i \in [0, m], j \in D_i],
\end{equation}
$D_i = \{ C_i \vee C_{\phi_i} \}$ denotes the pairwise training data for user $i$. $C_i$ is the observed items that user $i$ interacts with, while $C_{\phi_i}$ denotes the unobserved interactions. The second term in Eq. (\ref{equ: loss noise gene}) represents the Kullback-Leibler (KL) divergence between $f_\theta (N | \mathbf{z}_u, \mathbf{z}_i)$ and prior distribution $Q(N)$.
The distribution discrepancy loss function of KL divergence can be calculated as,
\begin{equation}
    \mathcal{L}_{ddl} = \mathcal{L}_{KLD} = \log \frac{1}{\sigma_{\theta}} + \frac{\sigma_\theta^2(\mathbf{z}_u, \mathbf{z}_i) + \mu_\theta(\mathbf{z}_u, \mathbf{z}_i)}{2} - \frac{1}{2}.
\label{equ: KLD}
\end{equation}

Furthermore, since the computational difficulty of directly calculating the KL divergence for non-standard normal distributions, we employ the non-parametric statistical method Maximum Mean Discrepancy (MMD) to quantify the discrepancy between the predefined distribution and the generated latent distribution,
\begin{equation}
\begin{split}
    \mathcal{L}_{ddl} =\mathcal{L}_{MMD} &= \mathbb{E}_{x,x^{'} \sim f_\theta}[k(x, x^{'})] + \mathbb{E}_{y,y^{'} \sim Q(N)}[k(y, y^{'})] \\
    &- 2\mathbb{E}_{x \sim f_\theta, y \sim Q(N)} [k(x,y)],
\end{split}
\label{equ: loss-mmd}
\end{equation}
where $k(\cdot,\cdot)$ denotes the kernel function. The first and second terms correspond to the expected value of the kernel function for pairs of samples drawn from the distribution $f_\theta$ and $Q(N)$, respectively. The third term is the expected value of the kernel function between samples drawn from $f_\theta$ and $Q(N)$, respectively. Combining the reconstruction loss with the distribution discrepancy loss, the data augmentation loss is presented as,
\begin{equation}
    \mathcal{L}_{aug} = \mathcal{L}_{recon} + \mathcal{L}_{ddl}.
\end{equation}

For data encoding, we adopt a well-known and widely recognized LightGCN \cite{he2020lightgcn} as the backbone. It is designed for recommendation and discards the feature transformation and nonlinear activation common in the Graph Convolutional Network (GCN) \cite{kipf2016semi}. The matrix form of the LightGCN convolution layer can be defined as,
\begin{subequations}
\begin{align}
    &\mathbf{z}^{(k)} = \tilde{A} \cdot \mathbf{\widetilde{z}}^{(k)} \label{equ: encoding1} \\
    &\mathbf{\widetilde{z}}^{(k)} = \mathbf{z}^{(k-1)} + f_{norm}(N^{(k)}), \label{equ: encoding2}
\end{align}
\end{subequations}
where $\mathbf{z}^{(k)} \in \{ \mathbf{z}_{ui}^{(k)} \in \mathbb{R} ^ {(m+n) \times d}, \mathbf{z}_{i}^{(k)} \in \mathbb{R} ^ {n \times d} \}$ is the embedding of convolutional layer $k$, $\mathbf{z}^{(0)} \in {\{E_{ui}, E_{i}\}}$ is the randomly initialized embedding. $N^{(k)} \in \{ N_{ui}^{(k)} \in \mathbb{R} ^ {(m+n) \times d}, N_{i}^{(k)} \in \mathbb{R} ^ {n \times d} \}$ is the generative noise. $\tilde{A} \in {\{A \in \mathbb{R} ^ {(m+n) \times (m+n)}, C \in \mathbb{R} ^ {n \times n}\}}$ and $\tilde{A}$ is the symmetrically normalized matrix. Equation (\ref{equ: encoding2}) serves to construct augmented views, corresponding to $f_g(\mathbf{z}_u, \mathbf{z}_i)$ in Eq. (\ref{equ: original_2}). The final representation we learned can be defined as follows,
\begin{equation} \label{final representation}
    \mathbf{z}=\frac{1}{L}\sum_{k=1}^{L}\mathbf{z}^{(k)},
\end{equation}
where $\mathbf{z} \in {\{\mathbf{z}_{u}, \mathbf{z}_{i}, \mathbf{z}_{c}\}}$ denotes the average representation and $L$ is the number of layers.

\subsubsection{Multi-Pair Contrast}
From the encoder, we obtain user embedding $\mathbf{z}_u$, item embedding $\mathbf{z}_i$, and complement embedding $\mathbf{z}_c$. As shown in Fig. \ref{framework}, we construct two contrastive views, the first layer embedding of encoder $\mathbf{z}^{'}=\mathbf{z}^{(1)}$, the average embedding of encoder $\mathbf{z}^{''}=\mathbf{z}$ through a single forward pass. The multi-pair contrastive loss $\mathcal{L}_{cl}$ is the average of three contrastive losses, which is defined as follows,
\begin{equation}
    \mathcal{L}_{cl} = avg(\mathcal{L}_{cl1} + \mathcal{L}_{cl2} + \mathcal{L}_{cl3}),
\end{equation}
where $avg(\cdot)$ is an average operation. $\mathcal{L}_{cl1}$, $\mathcal{L}_{cl2}$, and $\mathcal{L}_{cl3}$ correspond to user, item, and complement contrastive losses. As a commonly used contrastive learning loss function, InfoNCE\cite{oord2018representation} maximizes the mutual information between positive pairs and minimizes the mutual information between negative pairs. It is formulated as,
\begin{equation}
    \mathcal{L}_{cl_k} = \sum_{x\in \mathbb{B}}-\log
    \frac{\exp((z^{'}_{x} \cdot z^{''}_{x})/\tau)}{\sum_{y\in\mathbb{B}} \exp((z^{'}_{x}\cdot z^{''}_{y})/\tau)},
\end{equation}
where $k=\{1,2,3\}$, $\mathbb{B}$ is a mini-batch, $\exp(\cdot)$ is an exponential function, and $\tau$ is the temperature coefficient set to 0.2, which controls the similarity scale. The numerator $z^{'}_{x}$ and $z^{''}_{x}$ represent positive pairs whose consistency is reinforced, while the denominator $z^{'}_{x}$ and $z^{''}_{y}$ represent negative pairs, for which we reduce their consistency.

\subsubsection{Optimization Objectives}
For the primary recommendation task, we adopt the Bayesian Personalized Ranking (BPR) \cite{rendle2012bpr} loss function $\mathcal{L}_{rec}$,
\begin{equation}
\mathcal{L}_{rec} = \sum_{(u,i^{+},i^{-}) \in \mathbb{D}} -\log(\sigma(f_{u}(i^{+})-f_{u}(i^{-}))),
\end{equation}
\begin{equation}
    f_{u}(i) = \mathbf{z}_u \cdot \mathbf{z}_i,
\label{equ: prediction}
\end{equation}
where $\mathbb{D}$ denotes the dataset, $u$ and $i^{+}$ represent users and their positive item samples selected from the user-item interaction history, $i^{-}$ represents negative item samples randomly sampled from the entire dataset, $\log(\cdot)$ and $\sigma(\cdot)$ denote logarithm function and sigmoid function, $f_{u}(\cdot)$ denotes user's scores of positive or negative samples of items. The core idea behind BPR is that, given a user and two items, the model aims to learn the user's personalized preferences by ranking the more preferred item higher than the less preferred one.

We optimize the proposed model with a multi-task learning framework. The joint learning strategy integrates contrastive loss and data augmentation loss into the primary task to provide regularization for the recommendation. The overall loss function can be formalized as,
\begin{equation}
    \mathcal{L} = \mathcal{L}_{rec} + \lambda\mathcal{L}_{cl} + \mathcal{L}_{aug} + \mathcal{L}_{reg},
\end{equation}
where $\lambda$ is the coefficient for the contrastive loss, we set $\lambda=1$. $\mathcal{L}_{reg}$ is an $L_2$-norm regularization term that accounts for all parameters. It is worth noting that $\mathcal{L}_{aug}$ operates without fixed coefficients, as the noise scales are adaptively adjusted throughout the training process. Upon completion of the training process, the unknown preferences for recommendation are predicted using Eq. (\ref{equ: prediction}).

\subsection{Model Complexity}
We analyze the theoretical complexity of GDA4Rec within the scope of an iteration. Let $|E|$ and $\hat{E}$ be the number of non-zero elements in the user-item interaction matrix and the complement matrix, respectively, $d$ denote the embedding size, $L$ the number of encoding layers, $m$ and $n$ the numbers of users and items, $n^{+}$ and $n^{-}$ represent the counts of positive and negative item samples for each user. The space complexity is primarily driven by the data embeddings and model parameters. The data embedding consists of an adjacency matrix $A$, a complement matrix $C$, ego embeddings $E$, and noise $N$, with their respective complexities given as $\mathcal{O}((m+n)^2)$, $\mathcal{O}(n^2)$, $\mathcal{O}((m+n) \cdot d)$, and $\mathcal{O}((m+n) \cdot d)$. Since the reconstruction operation is applied only to a small number of samples, $\hat{R}$ is negligible. Similar to traditional embedding-based contrastive recommendations, model parameters are shared across all nodes. The storage space is very small and can be neglected. Regarding time complexity, the linear noise generation process has minimal impact on the overall complexity, which is primarily determined by the encoder and optimization modules. The encoder's time complexity is $\mathcal{O}(2|E| \cdot L \cdot d + |\hat{E}| \cdot L \cdot d)$, reflecting the convolutional operations performed on the user-item adjacency matrix and complement matrix. For optimization, we employ KL divergence in Eq. (\ref{equ: KLD}), with the complexity of $\mathcal{O}(d)$. The complexity of InfoNCE and BPR are $\mathcal{O}(m \cdot (d+n^{-}))$ and $\mathcal{O}(m \cdot n^{-} \cdot d)$, respectively.

\begin{table}[hp]
\small
\caption{Dataset statistics.}
\resizebox{\linewidth}{!}{
\begin{tabular}{ccccc}
\hline
Dataset   & User   & Item   & Interaction   &Density \\ \hline
CiaoDVD   & 17,615 & 16,121 & 72,665 & 0.025\%     \\
Yelp2     & 16,239 & 14,284 & 198,397 & 0.085\%   \\
Douban-book & 12,859  & 22,294  & 598,420 & 0.208\%     \\ \hline
\end{tabular}
\label{tab: datasets}
}
\end{table}

\section{Experiments}
\label{section: experiments}

\subsection{Experimental Settings}
\subsubsection{Datasets}
We use three real-world datasets, CiaoDVD, Yelp2, and Douban-book, to verify model performance with other state-of-the-art models.
We randomly sample 80\% interactions as training data, and the remaining 20\% as test data. The statistics of datasets are summarized in Table \ref{tab: datasets}, where density represents the proportion of actual interactions in the training set relative to all possible user-item pairs. Unlike many previous works, we conducted all experiments using 5-fold cross-validation to evaluate the model performance. Specifically, datasets are randomly divided into five equal-sized subsets. For each fold, the model is trained on four subsets and tested on the remaining one, ensuring that each subset serves as the test set once. The final performance metrics are averaged across all five folds to obtain a more robust estimate.

\subsubsection{Baselines}
To demonstrate the performance of our model, we compare GDA4Rec with state-of-the-art recommendation models in recent years.
The details of these models are as follows,
\begin{itemize}
\item $\textbf{LightGCN}$ (SIGIR, 2020) \cite{he2020lightgcn} abandons feature transformation and nonlinear activation in traditional GCN.

\item $\textbf{BUIR}$ (SIGIR, 2021) \cite{lee2021bootstrapping} adopts two distinct encoders that mutually learn from each other to mitigate the data sparsity. 

\item $\textbf{SGL}$ (SIGIR, 2021) \cite{wu2021self} leverages a GNN-based framework and contrastive learning to enhance node representations.

\item $\textbf{SSL4Rec}$ (CIKM, 2021) \cite{yao2021self} incorporates input information masking, dual-tower deep neural network encoding, and contrastive loss optimization to improve item representations.

\item $\textbf{NCL}$ (WWW, 2022) \cite{lin2022improving} incorporates a structure-contrastive objective and a prototype-contrastive objective.

\item $\textbf{SimGCL}$ (SIGIR, 2022) \cite{yu2022graph} replaces graph augmentations with uniform noise in the embedding space.

\item $\textbf{DAHNRec}$ (ESWA, 2024) \cite{zhu2024distribution} designs a distribution-aware noise augmentation method to enhance the uniformity of embeddings.

\item $\textbf{AUPlus}$ (ICWSM, 2024) \cite{ouyang2024improve} employs contrastive learning from the perspective of alignment and uniformity.
\end{itemize}

\begin{table*}[htbp]
\setlength{\tabcolsep}{3pt}
  \centering
  \caption{Performance comparison with other recommendation models.}
  \renewcommand{\arraystretch}{0.8}
    \begin{tabular}{ccccccccccccc}
    \toprule
    \multicolumn{1}{c}{Dataset} & Top@k & Metric & \makebox[1cm][c]{\fontsize{8}{8}\selectfont{LightGCN}} & \makebox[1cm][c]{\fontsize{8}{8}\selectfont{BUIR}}  & \makebox[1cm][c]{\fontsize{8}{8}\selectfont{SGL}}    & \makebox[1cm][c]{\fontsize{8}{8}\selectfont{SSL4Rec}} &  \makebox[1cm][c]{\fontsize{8}{8}\selectfont{NCL}}    & \makebox[1cm][c]{\fontsize{8}{8}\selectfont{SimGCL}} & \makebox[1cm][c]{\fontsize{8}{8}\selectfont{DAHNRec}} & \makebox[1cm][c]{\fontsize{8}{8}\selectfont{AUPlus}} & \makebox[1cm][c]{\fontsize{8}{8}\selectfont{GDA4Rec}} & \makebox[1cm][c]{\fontsize{8}{8}\selectfont{Improv.}} \\
    \midrule
    \midrule
    \multirow{9}[3]{*}{CiaoDVD} & \multirow{3}[2]{*}{5} & Precision & 0.0158  & 0.0148  & \underline{0.0182}  & 0.0112  & 0.0159  & 0.0172  & 0.0166  & 0.0167  & \textbf{0.0184 } & 1.10\% \\
          &       & Recall & 0.0447  & 0.0418  & \underline{0.0508}  & 0.0311  & 0.0457  & 0.0496  & 0.0477  & 0.0488  & \textbf{0.0528 } & 3.94\% \\
          &       & NDCG  & 0.0337  & 0.0312  & \underline{0.0380}  & 0.0239  & 0.0340  & 0.0372  & 0.0357  & 0.0364  & \textbf{0.0393 } & 3.42\% \\
          \cmidrule{2-13}          
          & \multirow{3}[1]{*}{10} & Precision & 0.0137  & 0.0130  & \underline{0.0150}  & 0.0091  & 0.0135  & 0.0143  & 0.0140  & 0.0130  & \textbf{0.0152 } & 1.33\% \\
          &       & Recall & 0.0747  & 0.0714  & \underline{0.0815}  & 0.0489  & 0.0749  & 0.0793  & 0.0783  & 0.0730  & \textbf{0.0846 } & 3.80\% \\
          &       & NDCG  & 0.0441  & 0.0415  & \underline{0.0484}  & 0.0299  & 0.0441  & 0.0475  & 0.0463  & 0.0448  & \textbf{0.0502 } & 3.72\% \\
          \cmidrule{2-13}
          & \multirow{3}[0]{*}{20} & Precision & 0.0110  & 0.0106  & \underline{0.0120}  & 0.0074  & 0.0112  & 0.0114  & 0.0115  & 0.0097  & \textbf{0.0123 } & 2.50\% \\
          &       & Recall & 0.1174  & 0.1144  & \underline{0.1258}  & 0.0761  & 0.1185  & 0.1219  & 0.1237  & 0.1045  & \textbf{0.1301 } & 3.42\% \\
          &       & NDCG  & 0.0562  & 0.0535  & \underline{0.0610}  & 0.0377  & 0.0565  & 0.0596  & 0.0592  & 0.0538  & \textbf{0.0631 } & 3.44\% \\
    \addlinespace[0.1cm]
    \hline
    \addlinespace[0.1cm]
   \multirow{9}[0]{*}{Yelp2} & \multirow{3}[0]{*}{5} & Precision & 0.0183  & 0.0142  & \underline{0.0206}  & 0.0138  & 0.0193  & 0.0203  & 0.0201  & 0.0101  & \textbf{0.0215 } & 4.37\% \\
          &       & Recall & 0.0316  & 0.0268  & \underline{0.0366}  & 0.0224  & 0.0338  & 0.0357  & 0.0353  & 0.0219  & \textbf{0.0384 } & 4.92\% \\
          &       & NDCG  & 0.0311  & 0.0249  & \underline{0.0351}  & 0.0219  & 0.0328  & 0.0343  & 0.0334  & 0.0197  & \textbf{0.0367 } & 4.56\% \\
          \cmidrule{2-13}
          & \multirow{3}[0]{*}{10} & Precision & 0.0152  & 0.0096  & 0.0171  & 0.0121  & 0.0162  & \underline{0.0173}  & 0.0171  & 0.0082  & \textbf{0.0185 } & 6.94\% \\
          &       & Recall & 0.0491  & \underline{0.0622}  & 0.0569  & 0.0370  & 0.0527  & 0.0571  & 0.0558  & 0.0331  & \textbf{0.0634 } & 1.93\% \\
          &       & NDCG  & 0.0365  & 0.0354  & \underline{0.0416}  & 0.0265  & 0.0389  & 0.0412  & 0.0400  & 0.0234  & \textbf{0.0441 } & 6.01\% \\
          \cmidrule{2-13}
          & \multirow{3}[0]{*}{20} & Precision & 0.0129  & 0.0096  & 0.0144  & 0.0103  & 0.0136  & \underline{0.0147}  & 0.0146  & 0.0071  & \textbf{0.0157 } & 6.80\% \\
          &       & Recall & 0.0792  & 0.0622  & \underline{0.0905}  & 0.0607  & 0.0849  & 0.0893  & 0.0896  & 0.0516  & \textbf{0.0962 } & 6.30\% \\
          &       & NDCG  & 0.0455  & 0.0354  & \underline{0.0516}  & 0.0336  & 0.0485  & 0.0510  & 0.0502  & 0.0290  & \textbf{0.0547 } & 6.01\% \\
    \addlinespace[0.1cm]
    \hline
    \addlinespace[0.1cm]
    \multirow{9}{*}{\parbox{1cm}{\centering Douban \newline book}} & \multirow{3}[0]{*}{5} & Precision & 0.0946  & 0.0526  & 0.1157  & 0.0831  & 0.1036  & \underline{0.1192}  & 0.1140  & 0.0974  & \textbf{0.1244 } & 4.36\% \\
          &       & Recall & 0.0687  & 0.0374  & \underline{0.0881}  & 0.0703  & 0.0791  & 0.0879  & 0.0853  & 0.0803  & \textbf{0.0917 } & 4.09\% \\
          &       & NDCG  & 0.1160  & 0.0635  & 0.1452  & 0.1070  & 0.1288  & \underline{0.1481}  & 0.1422  & 0.1187  & \textbf{0.1552 } & 4.79\% \\
          \cmidrule{2-13}
          & \multirow{3}[0]{*}{10} & Precision & 0.0763  & 0.0446  & 0.0913  & 0.0643  & 0.0832  & \underline{0.0944}  & 0.0901  & 0.0807  & \textbf{0.0982 } & 4.03\% \\
          &       & Recall & 0.1055  & 0.0626  & 0.1292  & 0.1005  & 0.1185  & \underline{0.1299}  & 0.1261  & 0.1199  & \textbf{0.1351 } & 4.00\% \\
          &       & NDCG  & 0.1169  & 0.0661  & 0.1448  & 0.1074  & 0.1302  & \underline{0.1473}  & 0.1416  & 0.1244  & \textbf{0.1540 } & 4.55\% \\
          \cmidrule{2-13}
          & \multirow{3}[1]{*}{20} & Precision & 0.0594  & 0.0365  & 0.0693  & 0.0478  & 0.0641  & \underline{0.0715}  & 0.0692  & 0.0627  & \textbf{0.0745 } & 4.20\% \\
          &       & Recall & 0.1552  & 0.0991  & 0.1816  & 0.1397  & 0.1701  & \underline{0.1849}  & 0.1809  & 0.1701  & \textbf{0.1912 } & 3.41\% \\
          &       & NDCG  & 0.1268  & 0.0746  & 0.1544  & 0.1154  & 0.1406  & \underline{0.1569}  & 0.1519  & 0.1367  & \textbf{0.1637 } & 4.33\% \\

    \bottomrule
    \end{tabular}%
  \label{tab: performance}%
\end{table*}%

\subsubsection{Settings}
 All the models are based on Pytorch, using Adam as the optimizer. For fair comparison, we set the same basic parameters for all models, where the embedding dimension is set to 64, the batch size is set to 2048, the learning rate is set to 0.001, and the regularization parameter is set to 0.0001 for better performance. We adopt a three-layer multilayer perceptron (MLP) as the generator $f_{\theta}$ and the reconstructor $f_{\phi}$. The number of convolutional layers for BUIR is set to 2 because it performs best with 2 layers, while the other models are set to 3. For other hyperparameters, we adopt the default configurations provided in the publicly available code. To address the class imbalance problem, we employ negative sampling. Specifically, for each positive user-item interaction, we randomly sample a fixed number of negative items that the user has not interacted with. Considering the ease of implementation for Gaussian distributions and KL divergence, we set the prior distribution $Q(N)$ as a Gaussian distribution and use KL divergence as the distribution discrepancy loss function, unless otherwise specified.

\subsubsection{Metrics}
We use several commonly adopted metrics to assess the performance: Precision, Recall, and Normalized Discounted Cumulative Gain (NDCG). For each metric, we compute values for the Top@5, Top@10, and Top@20 recommended items.

\subsection{Performance Evaluation}
We compare GDA4Rec with the baselines mentioned above. 
Based on the results in Table \ref{tab: performance}, we draw the following conclusions:

Compared to other baseline models, our model demonstrates superior performance across all datasets, particularly achieving over 6\% improvements in Top@20 metrics on the Yelp2 dataset. It is worth mentioning that on the sparsest dataset CiaoDVD (density = 0.025\%), GDA4Rec maintains about 3\% performance improvement over suboptimal model SGL. It also achieves significant improvements of 5.8\% and 6.7\% over SimGCL in terms of NDCG@20 and Recall@20, respectively, validating its effectiveness in handling sparse data scenarios.

With the exception of BUIR and SSL4Rec, all models adopt LightGCN as their backbone. Notably, these models consistently outperform the LightGCN across all datasets, except for AUPlus on CiaoDVD and Yelp2. This demonstrates the significant benefits of graph contrastive learning in improving performance. The performance drop of AUPlus may be attributed to the limited effectiveness of the alignment and uniformity losses on sparse datasets. 

Both BUIR and SSL4Rec exhibit inferior performance. We attribute this to the fact that BUIR focuses solely on the interactions between users and positive items, neglecting the information from potential negative samples. Additionally, SSL4Rec introduces significant alterations to the semantic information of the original graph through its random edge dropping and masking strategies, which results in performance degradation. This indicates that the organization of self-supervised signals plays a crucial role in performance improvement achieved by the model.

\subsection{Ablation Study}
In this section, we explore the extent to which each part of our model contributes to the final outcome. In response to this concern, we perform systematic ablation studies by comparing the GDA4Rec with other variants, and the quantitative results are demonstrated in Table \ref{tab: ablation} and Fig. \ref{fig: ablation}. The specific modifications for each variant are detailed as follows,
\begin{itemize}
\item  \textbf{w/o-cm}: 
This variant discards the complement matrix to evaluate the impact of item complementarity on model performance.

\item  \textbf{w/o-g}:
This variant removes the noise generation module from both data augmentation and encoding processes.

\item \textbf{w/o-f}: 
The function $f_{\mathrm{filter}}(\cdot)$ is a crucial step in generating the complement matrix. In this variant, we remove $f_{\mathrm{filter}}(\cdot)$.
\item \textbf{w/-rand}:
This variant substitutes the generative Gaussian noise with random Gaussian noise to validate the generative model.

\item \textbf{w/-un}: In the contrastive learning model, data augmentation is essential. Therefore, we perform a comparison variant by replacing Gaussian noise with uniform noise. To generate uniform noise, we use the Gaussian Error Linear Unit \cite{hendrycks2016gaussian} to transform generative Gaussian noise into uniform noise and further optimize it with Maximum Mean Discrepancy (MMD) to ensure better alignment with the uniform distribution.
\end{itemize}

\begin{table}[htbp] \footnotesize
\setlength{\tabcolsep}{2pt}
  \centering
  \caption{Performance on variants.}
  \renewcommand{\arraystretch}{0.8}
    \begin{tabular}{lcccccccc}
    \toprule
    \multicolumn{1}{c}{Dataset} & Top@k & Metric & \multicolumn{1}{c}{w/o-cm} & \multicolumn{1}{c}{w/o-g} & \multicolumn{1}{c}{w/o-f} & \multicolumn{1}{c}{w/-rand} & \multicolumn{1}{c}{w/-un} & GDA4Rec \\
    \midrule
    \midrule
    \multirow{9}[6]{*}{CiaoDVD} & \multirow{3}[2]{*}{5} & Precision & 0.0170  & 0.0167  & 0.0182 & 0.0169  & 0.0178  & \textbf{0.0184 } \\
          &       & Recall & 0.0488  & 0.0473  & 0.0515 & 0.0473 & 0.0500  & \textbf{0.0528 } \\
          &       & NDCG  & 0.0359  & 0.0358  & 0.0388 & 0.0355  & 0.0376  & \textbf{0.0393 } \\
\cmidrule{2-9}          & \multirow{3}[2]{*}{10} & Precision & 0.0146  & 0.0137  & 0.0151 & 0.0143  & 0.0146  & \textbf{0.0152 } \\
          &       & Recall & 0.0816  & 0.0749  & 0.0833 & 0.0792 & 0.0798  & \textbf{0.0846 } \\
          &       & NDCG  & 0.0474  & 0.0453  & 0.0497 & 0.0464 & 0.0477  & \textbf{0.0502 } \\
\cmidrule{2-9}          & \multirow{3}[2]{*}{20} & Precision & 0.0119  & 0.0112  & 0.0120 & 0.0117 & 0.0118  & \textbf{0.0123 } \\
          &       & Recall & 0.1274  & 0.1189  & 0.1290 & 0.1251 & 0.1245  & \textbf{0.1301}  \\
          &       & NDCG  & 0.0604  & 0.0577  & 0.0626 & 0.0593 & 0.0604  & \textbf{0.0631 } \\
    \midrule
    \multirow{9}[6]{*}{Yelp2} & \multirow{3}[2]{*}{5} & Precision & 0.0205  & 0.0202  & 0.0209 & 0.0211 & 0.0202  & \textbf{0.0215 } \\
          &       & Recall & 0.0363  & 0.0349  & 0.0366 & 0.0361 & 0.0357  & \textbf{0.0384 } \\
          &       & NDCG  & 0.0349  & 0.0342  & 0.0353 & 0.0349 & 0.0338  & \textbf{0.0367 } \\
\cmidrule{2-9}          & \multirow{3}[2]{*}{10} & Precision & 0.0174  & 0.0175  & 0.0178 & 0.0176 & 0.0173  & \textbf{0.0185 } \\
          &       & Recall & 0.0574  & 0.0562  & 0.0580 & 0.0565 & 0.0584  & \textbf{0.0634 } \\
          &       & NDCG  & 0.0417  & 0.0412  & 0.0422 & 0.0413 & 0.0413  & \textbf{0.0441 } \\
\cmidrule{2-9}          & \multirow{3}[2]{*}{20} & Precision & 0.0149  & 0.0148  & 0.0150 & 0.0150 & 0.0146  & \textbf{0.0157 } \\
          &       & Recall & 0.0926  & 0.0893  & 0.0921 & 0.0908 & 0.0935  & \textbf{0.0962 } \\
          &       & NDCG  & 0.0523  & 0.0511  & 0.0523 & 0.0516 & 0.0518  & \textbf{0.0547 } \\
    \midrule
    \multirow{9}[6]{*}{\parbox{1cm}{\centering Douban \newline book}} & \multirow{3}[2]{*}{5} & Precision & 0.1236  & 0.1166  & 0.1214 & 0.1189 & 0.1169  & \textbf{0.1244 } \\
          &       & Recall & 0.0903  & 0.0867  & 0.0878 & 0.0880 & 0.0884  & \textbf{0.0917 } \\
          &       & NDCG  & 0.1537  & 0.1445  & 0.1498 & 0.1480 & 0.1457  & \textbf{0.1552 } \\
\cmidrule{2-9}          & \multirow{3}[2]{*}{10} & Precision & 0.0978  & 0.0930  & 0.0957 & 0.0937 & 0.0932  & \textbf{0.0982}  \\
          &       & Recall & 0.1339  & 0.1287  & 0.1310 & 0.1295 & 0.1310  & \textbf{0.1351 } \\
          &       & NDCG  & 0.1526  & 0.1446  & 0.1491 & 0.1468 & 0.1461  & \textbf{0.1540 } \\
\cmidrule{2-9}          & \multirow{3}[2]{*}{20} & Precision & 0.0744  & 0.0710  & 0.0727 & 0.0715 & 0.0714  & \textbf{0.0745 } \\
          &       & Recall & 0.1898  & 0.1830  & 0.1863 & 0.1847 & 0.1866  & \textbf{0.1912 } \\
          &       & NDCG  & 0.1623  & 0.1545  & 0.1587 &  0.1567 & 0.1568  & \textbf{0.1637 } \\
    \bottomrule
    \end{tabular}%
  \label{tab: ablation}%
\end{table}%

\subsubsection{w/o-cm, w/o-g \& w/o-f}
\begin{figure}[tp]
\centering
\includegraphics[width=\linewidth]{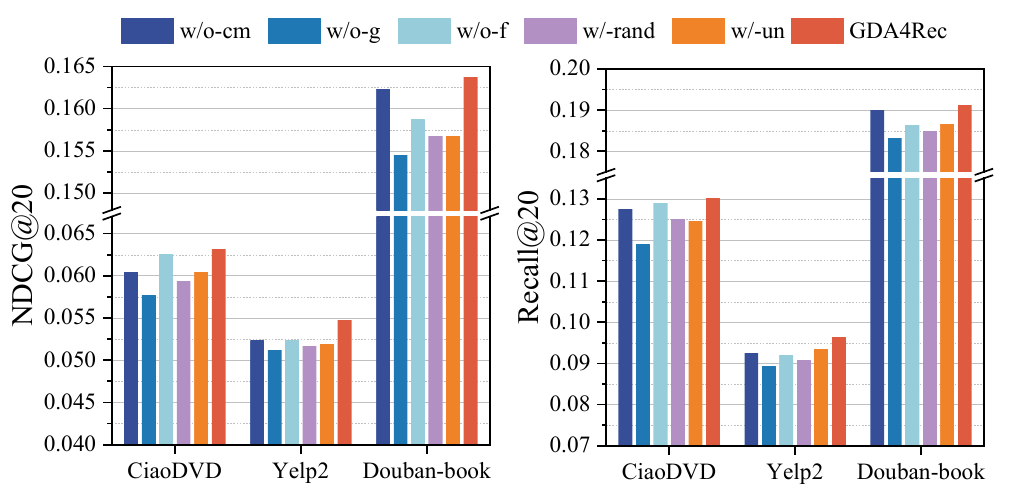}
\caption{Performance of different variants on NDCG@20 and Recall@20.}
\label{fig: ablation}
\end{figure}

The performance comparison presented in Table \ref{tab: ablation} and Fig. \ref{fig: ablation} indicates that removing the complement matrix and noise generation module leads to a decrease in model performance, particularly w/o-g. This demonstrates that the item complement matrix efficiently captures the latent correlation information between items, while generative noise produces high-quality augmented views. These modules collectively provide more precise supervision signals for the model. Moreover, not applying the function $f_{\mathrm{filter}}(\cdot)$ will also result in significant performance degradation. Actually, the inherent complexity of quantifying item correlations through user-item interactions often introduces unexpected noise into this process. The filtering mechanism helps mitigate this issue by disconnecting low-correlation items.

\subsubsection{w/-rand}
The experimental results show that GDA4Rec significantly outperforms w/-rand.
This demonstrates the superiority of our deep generative noise approach over conventional random noise methods in creating more meaningful and effective data augmentations. Furthermore, since GDA4Rec can adaptively adjust the parameter of noise generation during training, it simplifies the model's hyperparameter configuration.

\subsubsection{w/-un}
\begin{figure}[tp]
\centering
\includegraphics[width=\linewidth]{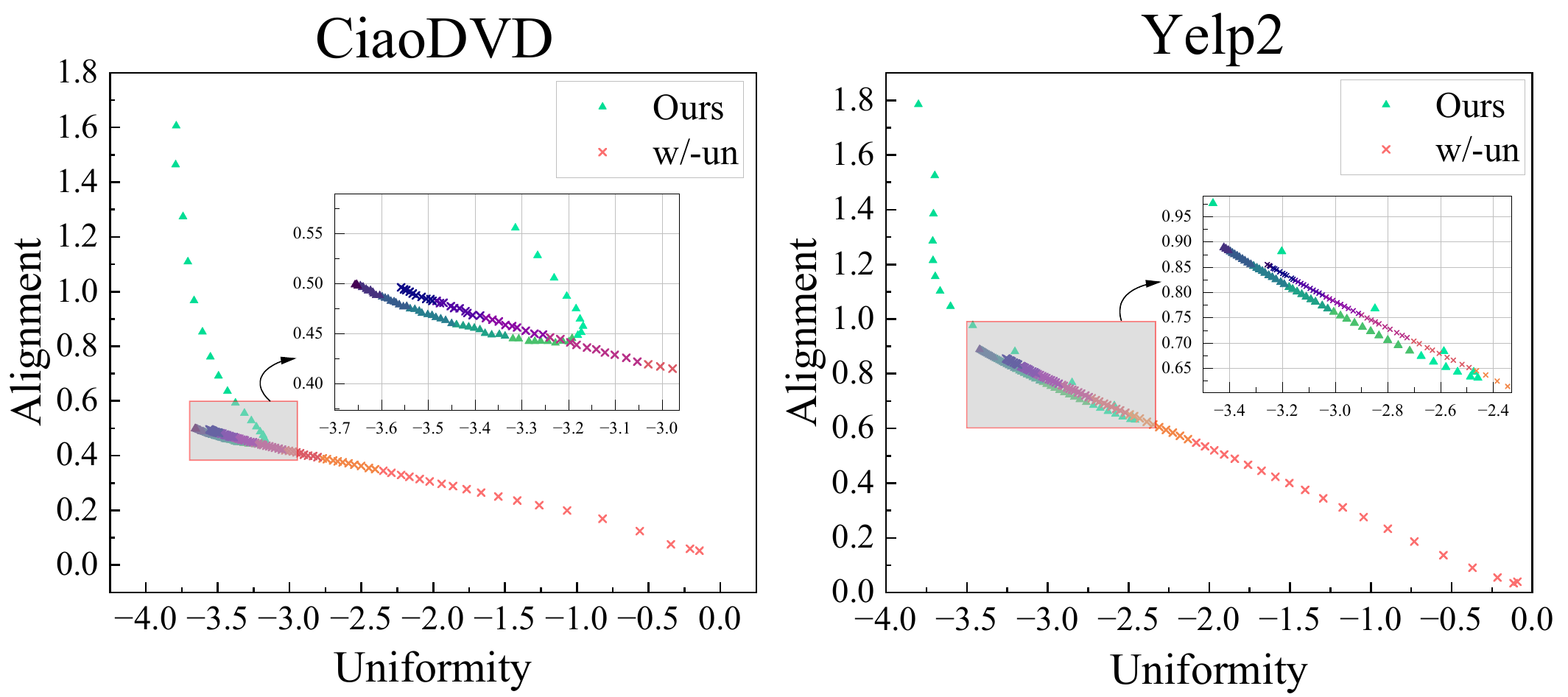}
\caption{The learning trajectories of alignment and uniformity on CiaoDVD and Yelp2. The color of the points ranges from light to dark, indicating the progression from the early to the later stages of training.}
\label{fig: CiaoDVD-yelp-AU}
\end{figure}

As shown in Table \ref{tab: ablation} and Fig. \ref{fig: ablation}, GDA4Rec outperforms $w/-un$, indicating that Gaussian noise has a significant positive effect on the model. We use two metrics \cite{wang2020understanding}, alignment and uniformity, to measure the impact of noise types on the model. Alignment measures the degree of approximation between positive samples, and uniformity measures the uniformity of the distribution of normalized features on unit hyperspheres. They are defined as,
\begin{equation}
\mathcal{L}_{align} = \mathbb{E}_{(z_{u},z_{i})\sim p_{u,pos}}||f_{\mathrm{norm}}(z_{u})-f_{\mathrm{norm}}(z_{i})||^{2},
\end{equation}
\begin{equation}
\mathcal{L}_{uniform} = \log\mathbb{E}_{z_{u}\sim p_{u}}e^{-2||z_{u}||^{2}}/2+
\log\mathbb{E}_{z_{i}\sim p_{pos}}e^{-2||z_{i}||^{2}}/2,
\end{equation}
where $p_{u}$ and $p_{pos}$ are the distribution of users and the distribution of positive pairs respectively. 
Figure \ref{fig: CiaoDVD-yelp-AU} illustrates the alignment and uniformity trajectories of w/-un and our model on the first-fold dataset. The figure reveals that during training, $w/-un$ tends to fit the data in a fixed direction, whereas GDA4Rec dynamically adjusts its optimization direction. This dynamic adaptability enables the model to explore a broader range of potential optimization distributions in the latent space, thereby enhancing its prediction performance significantly.

\begin{figure}[tp]
\centering
\includegraphics[width=\linewidth]{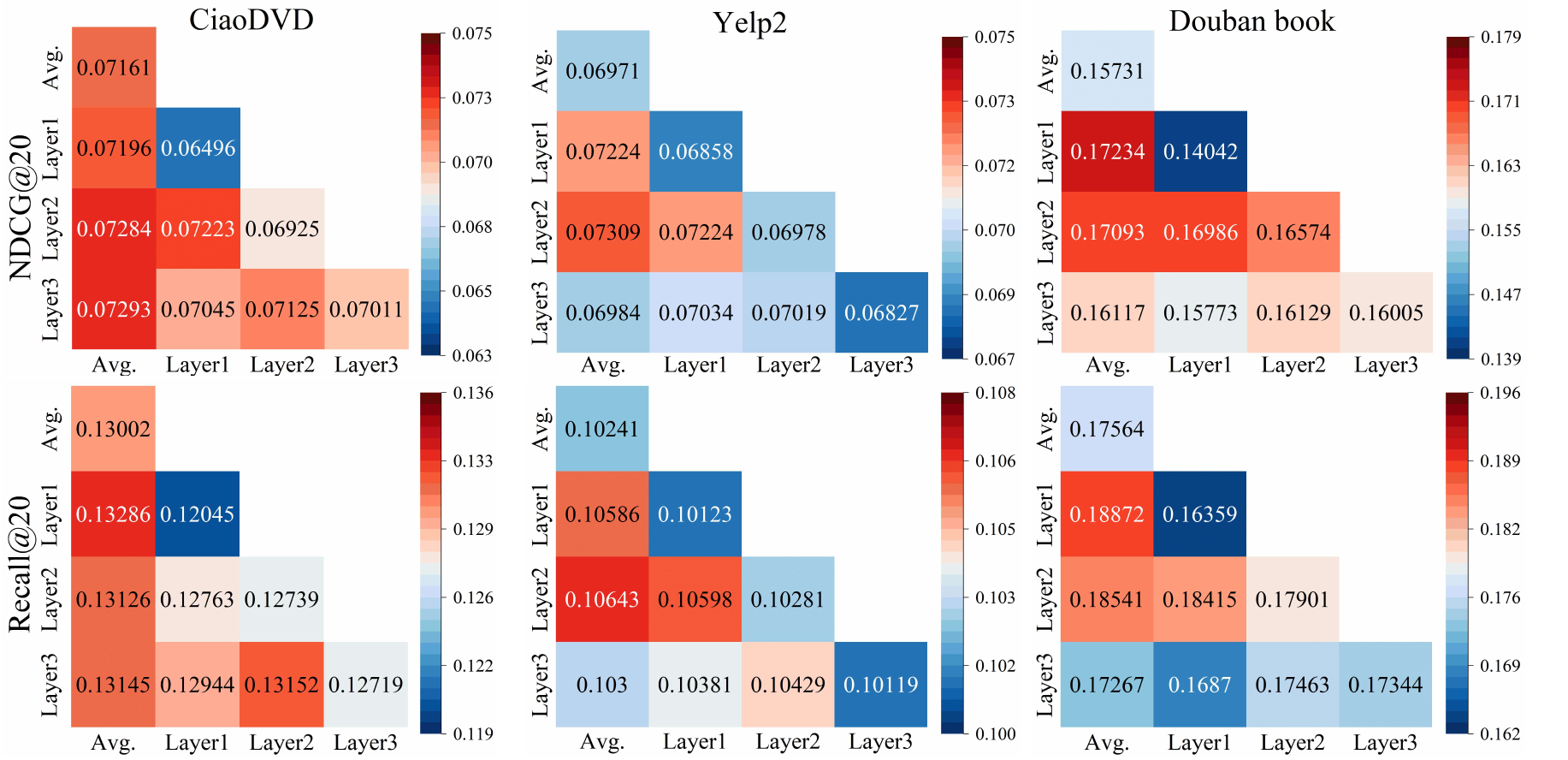}
\caption{Performance comparison of different contrastive layers in the first fold.}
\label{fig: heatmap}
\end{figure}

\begin{table}[htbp] \footnotesize
\setlength{\tabcolsep}{1.1pt}
  \centering
  \caption{Top@20 metrics for different encoder layers on CiaoDVD dataset.}
    \begin{tabular}{cccccccccc}
    \toprule
    \fontsize{6}{6}\selectfont {Layers} & \fontsize{6}{6}\selectfont{Metric} & \makebox[0.8cm][c]{\fontsize{6}{6}\selectfont{LightGCN}} & \makebox[0.8cm][c]{\fontsize{6}{6}\selectfont{SGL}}  & \makebox[0.8cm][c]{\fontsize{6}{6}\selectfont{NCL}}   & \makebox[0.8cm][c]{\fontsize{6}{6}\selectfont{SimGCL}} & \makebox[0.8cm][c]{\fontsize{6}{6}\selectfont{DAHNRec}} & \makebox[0.8cm][c]{\fontsize{6}{6}\selectfont{AUPlus}} & \makebox[0.8cm][c]{\fontsize{6}{6}\selectfont{GDA4Rec}} & \fontsize{6}{6}\selectfont{Improv.} \\
    \midrule
    \midrule
    \multirow{3}[2]{*}{2} & \fontsize{6}{6}\selectfont{Precision} & 0.0107  & 0.0114  & 0.0108  & 0.0102  & \underline{0.0115}  & 0.0098  & \textbf{0.0119 } & 3.48\% \\
          & \fontsize{6}{6}\selectfont{Recall} & 0.1134  & 0.1198  & 0.1142  & 0.1109  & \underline{0.1226}  & 0.1052  & \textbf{0.1254 } & 2.28\% \\
          & \fontsize{6}{6}\selectfont{NDCG}  & 0.0545  & \underline{0.0598}  & 0.0534  & 0.0539  & 0.0591  & 0.0548  & \textbf{0.0612 } & 2.34\% \\
    \midrule
    \multirow{3}[2]{*}{3} & \fontsize{6}{6}\selectfont{Precision} & 0.0110  & \underline{0.0120}  & 0.0112  & 0.0114  & 0.0115  & 0.0097  & \textbf{0.0123 } & 2.50\% \\
          & \fontsize{6}{6}\selectfont{Recall} & 0.1174  & \underline{0.1258}  & 0.1185  & 0.1219  & 0.1237  & 0.1045  & \textbf{0.1301 } & 3.42\% \\
          & \fontsize{6}{6}\selectfont{NDCG}  & 0.0562  & \underline{0.0610}  & 0.0565  & 0.0596  & 0.0592  & 0.0538  & \textbf{0.0631 } & 3.44\% \\
    \midrule
    \multirow{3}[2]{*}{4} & \fontsize{6}{6}\selectfont{Precision} & 0.0112  & \underline{0.0117}  & 0.0114  & 0.0109  & 0.0115  & 0.0075  & \textbf{0.0124 } & 5.98\% \\
          & \fontsize{6}{6}\selectfont{Recall} & 0.1196  & 0.1219  & 0.1204  & 0.1175  & \underline{0.1240}  & 0.0819  & \textbf{0.1312 } & 5.81\% \\
          & \fontsize{6}{6}\selectfont{NDCG}  & 0.0566  & \underline{0.0613}  & 0.0578  & 0.0579  & 0.0590  & 0.0441  & \textbf{0.0638 } & 4.08\% \\
    \midrule
    \multirow{3}[2]{*}{5} & \fontsize{6}{6}\selectfont{Precision} & 0.0111  & 0.0116  & 0.0112  & 0.0113  & \underline{0.0117}  & 0.0063  & \textbf{0.0125 } & 6.84\% \\
          & \fontsize{6}{6}\selectfont{Recall} & 0.1186  & 0.1209  & 0.1190  & 0.1195  & \underline{0.1254}  & 0.0700  & \textbf{0.1325 } & 5.66\% \\
          & \fontsize{6}{6}\selectfont{NDCG}  & 0.0559  & \underline{0.0607}  & 0.0567  & 0.0596  & 0.0604  & 0.0374  & \textbf{0.0646 } & 6.43\% \\
    \bottomrule
    \end{tabular}%
  \label{tab: encoder_layers}%
\end{table}%

\subsection{Contrastive Layers for Model Performance}
To thoroughly evaluate the effect of contrastive learning, we compute the contrastive loss between all possible combinations of layers. Figure \ref{fig: heatmap} illustrates the NDCG@20 for the first fold of the cross-validation dataset. The figure reveals that the optimal contrastive layer combinations vary across different datasets, but the most effective ones typically involve the average layer in conjunction with other layers. This observation allows us to concentrate on combinations involving the average layer with other layers, dramatically simplifying the contrastive layer selection process. As shown in the figure, the contrast between the average layer and the first layer consistently yields satisfactory results. Thus this combination is chosen for contrastive loss calculation in our model.

\begin{figure}
\centering
\includegraphics[width=0.8\linewidth]{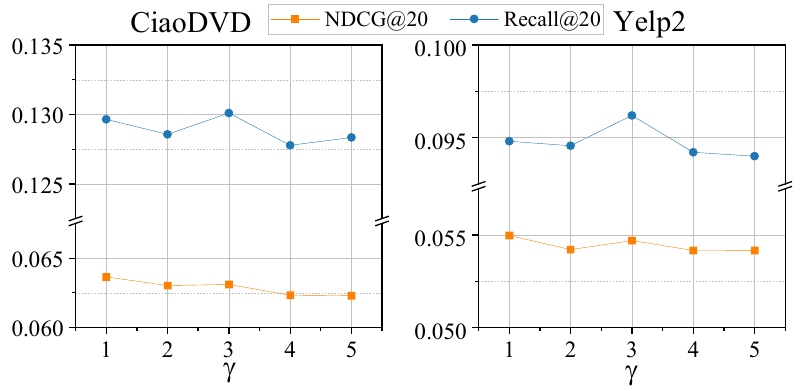}
\caption{The performance of different $\gamma$ on CiaoDVD and Yelp2.}
\label{fig: hypers}
\end{figure}

\subsection{Hyperparameter Analysis}
Our framework utilizes a deep generative model for data augmentation to produce contrastive views. This innovative architecture achieves adaptive parameter adjustment capabilities, eliminating the requirement for manual tuning of hyperparameters such as noise scale and coefficient of the data augmentation loss. In this part, we analyze two key hyperparameters, the number of layers $L$ in the encoder and the filtering threshold $\gamma$ in the filter function. The impact of these parameters on the model is evaluated by Top@20 metrics, shown in Table \ref{tab: encoder_layers} and Fig. \ref{fig: hypers}.

It is observed that the number of encoder layers, $L$, has the most significant impact on model performance from Table \ref{tab: encoder_layers}. As the number of layers increases, the model is capable of capturing more complex graph structural information. However, an excessive number of layers may lead to over-smoothing, where node representations become too similar, thereby degrading performance. Additionally, increasing the number of layers incurs higher computational costs. As illustrated in Table \ref{tab: encoder_layers},
SGL, SimGCL, and AUPlus, which generate augmented views using random methods, exhibit varying degrees of performance degradation as the number of layers increases. This suggests that while such augmentation introduces view diversity, it may alter the original semantic structure, making the model more susceptible to over-smoothing. In contrast, NCL and DAHNRec preserve the graph's original characteristics through neighbor information and linear networks respectively, effectively mitigating this risk. GDA4Rec, utilizing a generative model for data augmentation, maintains semantic integrity while introducing controlled diversity.
In our experiments, we set $L=3$ as the number of encoder layers for fair comparison and computational efficiency. But GDA4Rec possesses significant potential for further performance improvements with more layers.

The parameter $\gamma$ serves as a threshold, where any relationship strength in the item complement matrix below this value is set to zero. This operation is designed to filter out weakly associated item pairs. By adjusting $\gamma$, the model can focus on more significant item relationships while transforming the matrix into a sparse one to improve computational efficiency. As shown in Fig. \ref{fig: hypers}, the model performance exhibits best when $\gamma$ is equal to 3. Thus we set $\gamma=3$ considering both model performance and computational efficiency.

\section{Conclusion and Future Work}
\label{section: conclusion}
Recommendation systems have been a well-established research topic, recent efforts have increasingly centered around leveraging contrastive learning methods to overcome the challenges posed by data sparsity.
In this paper, we propose a novel framework, namely GDA4Rec, which incorporates deep generative models to construct augmented views. Through reconstruction loss and distribution discrepancy loss, these augmented views not only effectively preserve the original semantic information of users and items but also enhance the data diversity. Additionally, we extract an item complement matrix from the user-item interaction matrix to capture the latent correlations between items. By integrating contrastive learning strategies, these methods provide rich self-supervised signals to the model, effectively alleviating the impact of data sparsity and thereby improving recommendation performance. In the experiment, we thoroughly analyze the effectiveness of each module and provide insights into their underlying mechanisms. 

The results presented in this paper demonstrate the promise of incorporating the deep generative model into contrastive learning. Future work should focus on further enhancing the scalability and efficiency of the model. Additionally, exploring the potential applications of this method in other domains presents a meaningful direction, with the potential to unlock new insights and solutions.

\begin{acks}
This work is supported by the Natural Science Foundation of China (No. 62402398, No. 72374173), the Natural Science Foundation of Chongqing (No. CSTB2025NSCQ-GPX1082), the University Innovation Research Group of Chongqing (No. CXQT21005), the Chongqing Graduate Research and Innovation Project (No. CYB23124), the Fundamental Research Funds for the Central Universities (No. SWU-KR24025, No. SWU-XDJH202303) and the High Performance Computing clusters at Southwest University.
\end{acks}

\clearpage
\section{GenAI Usage Disclosure}
During the preparation of this work, the author(s) utilized generative AI tools including ChatGPT, DeepSeek, and Grok to assist with language refinement and editing. After using these tools/services, the author(s) carefully reviewed and edited the content as needed and take(s) full responsibility for the content of the publication.

\bibliographystyle{ACM-Reference-Format}
\balance
\bibliography{ref}

\end{document}